\documentclass[aps,prl,twocolumn,superscriptaddress,showpacs,floatfix]{revtex4}

\usepackage{amsmath,amssymb,bm}
\usepackage{graphicx}

\begin{document}

\title{Energy Gaps in Graphene Nanoribbons}
\author{Young-Woo Son}
\author{Marvin L. Cohen}
\author{Steven G. Louie}
\email[Email:\ ]{sglouie@berkeley.edu}
\affiliation{Department of Physics, 
             University of California at Berkeley, 
             Berkeley, CA 94720}
\affiliation{Materials Sciences Division, 
             Lawrence Berkeley National Laboratory, 
             Berkeley, CA 94720}
\date{\today }
\begin{abstract}
Based on a first-principles approach, 
we present scaling rules for the band gaps of
graphene nanoribbons (GNRs) as a function of their widths.
The GNRs considered have either armchair or
zigzag shaped edges on both sides 
with hydrogen passivation.
Both varieties of ribbons 
are shown to have band gaps.
This differs from the results of simple tight-binding 
calculations or solutions of the Dirac's equation based on them. 
Our {\it ab initio} calculations show that 
the origin of energy gaps for GNRs 
with armchair shaped edges arises from both quantum
confinement and the crucial effect of the edges.
For GNRs with zigzag shaped edges,
gaps appear because of a staggered sublattice potential 
on the hexagonal lattice
due to edge magnetization.
The rich gap structure for ribbons with 
armchair shaped edges
is further obtained analytically including edge effects. 
These results reproduce our {\it ab initio} calculation results very well. 
\end{abstract}

%73.22.-f Electronic structure of nanoscale materials: 
%         clusters, nanoparticles, nanotubes, and nanocrystals 
%75.70.Ak Magnetic properties of monolayers and thin films
%72.80.Rj Fullerenes and related materials 
\maketitle
The electronic structure of nanoscale carbon materials
such as fullerenes and carbon nanotubes
has been the subject of intensive research during
the past two decades~\cite{dressel}
because of fundamental scientific interest in nanomaterials
and because of their versatile electronic properties
that are expected to be important for future nanoelectronics~\cite{mceuen,louie}.
Among the carbon nanostructures, a simple variation of graphene, 
ribbons with nanometer sized widths, has been studied 
extensively~\cite{fujita,nakada,waka,ezawa,brey,abanin,sasaki,okada,hslee,half,miyamoto,kawai}.
Because of recent progress in preparing 
single graphite layers on conventional device setups, 
the graphene nanoribbons (GNRs) with varying widths 
can be realized either by cutting~\cite{hiura} 
mechanically exfoliated graphenes~\cite{novoselov,kim} 
or by patterning epitaxially grown graphenes~\cite{berger,berger2}.

Since GNRs are just geometrically terminated single graphite layers, 
their electronic structures have been modeled by imposing appropriate
boundary conditions on 
Schr\"odinger's equation with simple tight-binding (TB)
approximations based on $\pi$-states of carbon~\cite{fujita,nakada,waka,ezawa} 
or on a 2-dimensional free massless 
particle Dirac's equation with an effective
speed of light ($\sim 10^6$m/s)~\cite{brey,abanin,sasaki}.
Within these models, it is predicted that 
GNRs with armchair shaped edges can be either metallic or semiconducting 
depending on their widths~\cite{fujita,nakada,waka,ezawa,brey,abanin,sasaki},
and that GNRs with zigzag shaped edges are
metallic with peculiar edge states on 
both sides of the ribbon regardless of its
widths~\cite{fujita,nakada,waka,ezawa,brey,abanin,sasaki,okada,hslee,half,miyamoto}.

Although the aformentioned models are known 
to describe the low energy properties of graphene 
very well~\cite{novoselov,kim,berger,berger2,sadowski,reich},
a careful consideration of edge effects 
in nanometer sized ribbons
are required to determine their bandgaps
because, unlike the situation in graphene, 
the bonding characteristics between atoms change
abruptly at the edges~\cite{ezawa,kawai}.
Moreover, the spin degree of freedom 
is also important because the GNRs with zigzag shaped edges
have narrow-band edge states at the Fermi energy ($E_F$)
implying possible magnetization at the edges~\cite{fujita,okada,hslee,half}. 
Motivated by the recent experimental progress in this area,
we have carried out first-principles calculation and theoretical 
analysis to explore the relation between the bandgap 
and the geometries of GNRs.

In this Letter, we show that GNRs with hydrogen passivated armchair
or zigzag shaped edges both always have nonzero and direct bandgaps. 
The origins of the bandgaps for the 
different types of homogenous edges vary.
The bandgaps of GNRs with armchair shaped edges originate from
quantum confinement, and edge effects play a crucial role. 
For GNRs with zigzag shaped edges, 
the bandgaps arise from a staggered sublattice potential due to 
spin ordered states at the edges~\cite{fujita,okada,hslee,half}.
Although the ribbon widths and energy bandgaps of the GNRs are 
related to each other primarily in inverse proportion, 
there is a rich structure 
in the ratio of the proportionalities as in the behavior of 
carbon nanotubes~\cite{dressel}. 
For GNRs with armchair edges, analytic scaling rules
for the size of the bandgaps 
are obtained as a function of width 
including the effect of the edges and give
a good agreement with our first-principles calculations.
%For GNRs with armchair edges, we obtain analytic scaling rules
%for the size of the bandgaps as a function of width when we 
%consider the effect of the edges. 
%These results from the analytical expressions 
%are in a good agreement 
%with our first-principles calculations.

Our electronic structure calculation employs the 
first-principles self-consistent pseudopotential
method~\cite{soler} using the local (spin)
density approximation (L(S)DA)~\cite{soler,gga}.
An energy cutoff (for a real space mesh size) 
of 400 Rydbergs is employed and 
a double-$\zeta$ plus polarization basis set~\cite{soler} 
is used for the localized basis orbitals to deal with
the many atoms in a unit cell of the GNRs of various widths.
We obtained the electron density by integrating the density 
matrix with a Fermi-Dirac distribution~\cite{soler,fd}. 
The geometry for each GNR studied is fully relaxed 
until the force on each of the atoms is less than 16pN.
A k-point sampling of 32 (96) k-points that are uniformly positioned 
along the 1D Brillouin zone is employed
for GNRs with armchair (zigzag) shaped edges. 
We have tested the change of gap size by increasing 
the vacuum between edges from 20 to 40 \AA~and 
between plances from 16 to 25 \AA~and found no changes.

\begin{figure}[t]
\centering
\includegraphics[width=7.5cm]{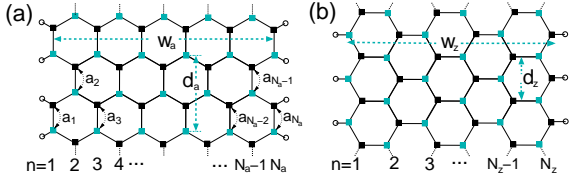}
\caption{
(a) Schematic of a 11-AGNR.
The empty circles denote hydrogen atoms
passivating the edge carbon atoms, and 
the black and grey rectangles represent
atomic sites belonging to different sublattice 
in the graphene structure.
The 1D unit cell distance and ribbon width 
are represented by $d_a$ and $w_a$ respectively.
The carbon-carbon distances on the $n$-th dimer line
is denoted by $a_n$.
(b) Schematic of a 6-ZGNR.
The empty circles and rectangles follow the same
convention described in (a).
The 1D unit cell distance and the ribbon width 
are denoted by $d_z$ and $w_z$ respectively.
\vspace{-4.0mm}
}
\end{figure}

Following previous 
convention~\cite{fujita,nakada,waka,ezawa,brey,abanin,sasaki,okada,hslee,half,miyamoto,kawai},
the GNRs with armchair shaped edges on both sides
are classified by the number of dimer lines ($N_a$)
across the ribbon width as shown in Fig. 1(a). 
Likewise, ribbons with zigzag shaped edges on both sides 
are classified by the number of 
the zigzag chains ($N_z$) across the ribbon width [Fig. 1(b)].
We refer to a GNR with $N_a$ dimer lines 
as a $N_a$-AGNR and a GNR with $N_z$ zigzag chains
as a $N_z$-ZGNR.

Our calculations show that the $N_a$-AGNRs
are semiconductors with energy gaps which 
decrease as a function of increasing ribbon widths ($w_a$).
The variations in energy gap however exhibit three 
distinct family behaviors [Fig. 2].
Moreover, the energy gaps obtained by a simple TB model
are quite different from those by first-principles calculations.
The TB results using a constant
nearest neighbor hopping integral, $t=2.7$ eV~\cite{reich}
between $\pi$-electrons 
are summarized as function of width in Fig. 2(a).
It shows that a $N_a$-AGNR is metallic 
if $N_a=3p+2$ (where $p$ is a positive integer)
or otherwise, it is semiconducting,
in agreement with previous 
calculations~\cite{fujita,nakada,waka,ezawa,brey,sasaki,abanin}.
The gap of a $N_a$-AGNR ($\Delta_{N_a}$)
is inversely proportional to its width,
separated into basically two groups 
with a hierarchy of gap size given by
$\Delta_{3p}\gtrsim\Delta_{3p+1}>\Delta_{3p+2}(=0)$ for all $p$'s.
For the first-principles calculations, however,
there are {\it no} metallic nanoribbons.
The gaps as a function of ribbon width are now well
separated into three different 
categories (or family structures) as shown in Fig. 2(b).
Moreover, the gap size hierarchy is also changed to 
$\Delta_{3p+1}>\Delta_{3p}>\Delta_{3p+2}(\neq0)$.
For example, in the first-principles calculation for $p=13$,
the lowest energy gap is $\Delta_{38}=45$ meV 
and $\Delta_{40}-\Delta_{39}= 68$ meV, all of 
which are quite larger values compared to those (0 and $-2$ meV respectively)
obtained from TB approximations. 
The first-principles band structures of $N_a$-AGNRs are shown in 
Fig. 2(c) for $N_a$=12, 13, and 14. They exhibit direct
bandgaps at $kd_a=0$ for all cases.

\begin{figure}[t]
\centering
\includegraphics[width=7.5cm]{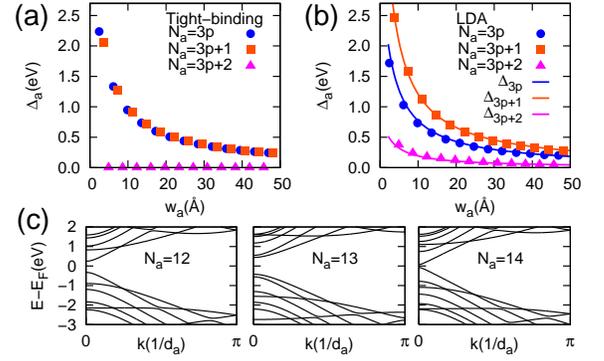}
\caption{(color online)
The variation of bandgaps of $N_a$-AGNRs
as a function of width ($w_a$) obtained 
(a) from TB calculations with  $t=2.70$ (eV) and
(b) from first-principles calculations (symbols). 
The solid lines in (b) are from Eq. (1).
(c) First-principles band structures of $N_a$-AGNRs with $N_a$=12, 13, and 14,
respectively.
\vspace{-4.0mm}
}
\end{figure}

A determining factor in the semiconducting behavior of $N_a$-AGNR
is quantum confinement which can be characterized by 
$\Delta_{N_a}\sim w_a^{-1}$~\cite{fujita,nakada,waka,ezawa,brey,sasaki,abanin}. 
In addition, as will be discussed below, 
the edge effects play a crucial role and force the
$(3p+2)$-AGNRs (predicted to be metallic by TB model)
to be semiconductors.
The edge carbon atoms of our AGNRs are passivated by
hydrogen atoms (by some foreign atoms/molecules in general) 
so that the $\sigma$ bonds between hydrogen
and carbon and the onsite energies of the carbons at the edges
would be different from those in the middle of the AGNRs.
The bonding distances between carbon atoms 
at the edges are also expected to change accordingly.
Such effects have been observed in large aromatic molecules,
e.g., ovalene (C$_{32}$H$_{14}$)~\cite{coulson}.
In Fig. 3(a), we show that the bond lengths parallel to dimer lines
at edges ($a_1$ and $a_{N_a}$ for $N_a$-AGNR in Fig. 1 (a)) 
are shortened by 3.3$\sim$3.5\% for the 12-, 13-, and 14-AGNR
as compared to those in the middle of the ribbon. 
From the analytic expressions for TB matrix elements between
carbon atoms in Ref. 27, 
a 3.5\% decrease in interatomic distance 
from 1.422\AA~would induce a 12\% increase in the hopping integral
between $\pi$-orbitals.

\begin{figure}[t]
\centering
\includegraphics[width=8.0cm]{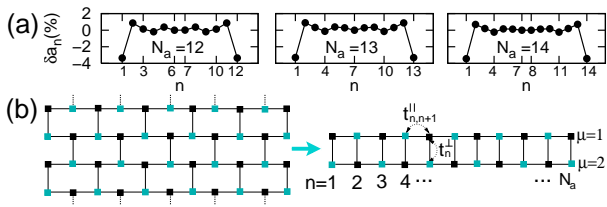}
\caption{
(a) The ratio of the calculated change in the carbon-carbon 
distance ($\delta a_n$) (see Fig. 1(a)) to 
the carbon-carbon distance in the
middle of the $N_a$-AGNRs, i.e.,
$\delta a_n\equiv100\times\frac{a_n-a_c}{a_c}$
where $a_c=a_6=a_7=1.424$ \AA~for $N_a=12$, 
$a_c=a_7=1.422$ \AA~for $N_a=13$,
and $a_c=a_7=a_8=1.423$ \AA~for $N_a=14$, respectively.
(b) Topologically equivalent structure to the $N_a$-AGNR shown in Fig. 1(a).
For the especial case of $k=0$, a lattice with periodic 
ladders (left) can be folded into a
two-leg ladder with $N_a$ rungs (right).
\vspace{-4.0mm}
}
\end{figure}

To see the consequence of such effects more clearly, 
we introduce a lattice model [Fig. 3(b)] which is 
equivalent to the AGNRs within the TB approximation~\cite{fujita,nakada,waka,ezawa}. 
The set of eigenvalues of a brick type lattice shown in Fig 3.(b) at $kd_a=0$ 
is further equivalent to that of a two-leg ladder system 
with $N_a$ rungs~\cite{fujita,nakada,waka,ezawa}.
The Hamiltonian of this simpler model can be written as 
%\begin{eqnarray}
%{\mathcal H} &=& 
%\sum_{n=1}^{N_a}\sum_{\mu=1}^2
%\varepsilon_{\mu,n}a^\dagger_{\mu,n} a_{\mu,n}
%-\sum_{n=1}^{N_a} t^\perp_n
%\left( a^\dagger_{1,n}a_{2,n}+\text{h.c.}\right)\nonumber\\
%& &-\sum_{n=1}^{N_a-1}\sum_{\mu=1}^2 
%t^{||}_{n,n+1}\left(a^\dagger_{\mu,n} a_{\mu,n+1}+\text{h.c.}\right),
%\end{eqnarray}
${\mathcal H}=
\sum_{n=1}^{N_a}\sum_{\mu=1}^2
\varepsilon_{\mu,n}a^\dagger_{\mu,n} a_{\mu,n}
-\sum_{n=1}^{N_a} t^\perp_n
(a^\dagger_{1,n}a_{2,n}+\text{h.c.})
-\sum_{n=1}^{N_a-1}\sum_{\mu=1}^2 
t^{||}_{n,n+1}(a^\dagger_{\mu,n} a_{\mu,n+1}+\text{h.c.}),
$
where \{$n$, $\mu$\} denote a site,
$\varepsilon_{\mu,n}$ site energies,
$t^{||}_{n,n+1}$ and $t^\perp_n$ the nearest neighbor hopping integrals
within each leg and between the legs respectively, 
and $a_{\mu,n}$ the annihilation operator of $\pi$-electrons
on the $n$-th site of the $\mu$-th leg.
As discussed above and shown in Fig. 3(a), 
$t^\perp_n$ and $\varepsilon_{\mu,n}$ at the edges would
differ from those in the middle of GNRs.  
Hence, considering the simplest but essential variation from the exact solvable model
to approximate the realistic situations, 
we assume that $t^\perp_1=t^\perp_{N_a}\equiv(1+\delta)t$ and
$t^\perp_n\equiv t$  for $n=2,\cdots,N_a-1$, and 
$t^{||}_{n,n+1}\equiv t$ for all $n$'s.
The site energies are set at 
$\varepsilon_{\mu,n}\equiv\varepsilon_0$ for $n=1$ and $N_a$
and 0 otherwise regardless of $\mu$.
This model Hamiltonian is solved pertubatively and
the resulting energy gaps to the first order in
$\delta$ and $\varepsilon_0$ are as follows,
\begin{eqnarray}
\Delta_{3p} &\simeq&
\Delta^0_{3p}
-\frac{8\delta t}{3p+1}\sin^2\frac{p\pi}{3p+1},\nonumber\\
\Delta_{3p+1} &\simeq& 
\Delta^0_{3p+1}
+\frac{8\delta t}{3p+2}\sin^2\frac{(p+1)\pi}{3p+2},\nonumber\\
\Delta_{3p+2}&\simeq&\Delta^0_{3p+2}+\frac{2|\delta| t}{p+1},
\end{eqnarray}
where $\Delta_{3p}^0$, $\Delta_{3p+1}^0$ and $\Delta_{3p+2}^0$
are the gaps of the ideally terminated ribbon 
when $\delta=\varepsilon_0=0$. 
They are given by
$t\left[4\cos\frac{p\pi}{3p+1}-2\right]$,
$t\left[2-4\cos\frac{(p+1)\pi}{3p+2}\right]$ and 0 respectively.
The zeroth-order gaps are identical to the values
obtained from numerical calculations 
in Fig. 2(a)~\cite{fujita,nakada,waka,ezawa}.
With $t=2.7$ (eV) ~\cite{reich} and $\delta=0.12$, 
the calculated gaps obtained using Eq. (1) 
are in good agreement with our LDA results [Fig. 2(b)].
This implies that the 12\% increase of the hopping 
integrals between carbon atoms at the edges opens the 
gaps of the $(3p+2)$-AGNRs and decreases (increases) 
the gaps of $3p$-AGNRs~($(3p+1)$-AGNRs). 
This analysis provide the physical explanation of the changes
in the gap hierarchy discussed before.
We note that there is 
no contribution from the variation 
in the site energies ($\varepsilon_0$) 
at the edges to first order.

Next, we find that nanoribbons with zigzag shaped edges 
also have direct bandgaps which decrease
with increasing width ($w_z$).
The eigenstates of the ZGNRs near $E_F$, without
considering spins, have a peculiar edge-state structure.
As noted earlier within the tight-binding picture~\cite{fujita}, 
there are two edge states decaying into the center of the ZGNR with
a decay profile depending on their momentum
as $\sim e^{-\alpha_k r}$ where 
$\alpha_k\equiv-\frac{2}{\sqrt{3}d_z}\ln\left|2\cos\frac{kd_z}{2}\right|$ 
($\frac{2\pi}{3}\le kd_z \le \pi$, $d_z=$ unit cell length shown in Fig. 1(b)).
Our first-principles calculation also predicts a set of doubly degenerate flat 
edge-state bands at $E_F$ when not considering spins (not shown here).
Since the edge-states around $E_F$ form flat bands,
they give rise to a very large density of states at $E_F$.
Thus infinitesimally small onsite repulsions 
could make the ZGNRs magnetic~\cite{fujita},
unlike the case with two dimensional graphene which has a 
zero density of states at $E_F$.
As pointed out in a TB study earlier~\cite{fujita}
and later confirmed by first-principles studies~\cite{okada,hslee,half}, 
our LSDA calculation also shows 
that the ground state of ZGNRs with hydrogen passivated
zigzag edges indeed have finite magnetic moments on each edge
with negligible change in atomic structure~\cite{okada,hslee,half,miyamoto}.

\begin{figure}[t]
\centering
\includegraphics[width=7.5cm]{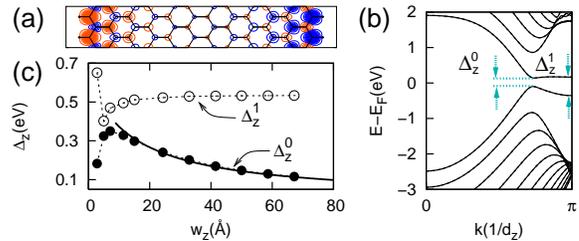}
\caption{(color online)
(a) 
Contour graph for 
$\rho_{\alpha}(r)-\rho_{\beta}(r)$
of a 12-ZGNR (the density
is integrated over the normal direction to the ribbon plane). 
The lowest (highest) contour  
of $\pm0.4\times10^{-4}a^{-2}_0$ 
is drawn by a thick blue (red) line 
and the spacing for blue (red) lines is
$1.0\times10^{-4}a^{-2}_0$ 
($a_0$ = Bohr radius).
(b) 
The band structure of a 12-ZGNR.
The $\alpha$- and $\beta$-spin states
are degenerate in all energy bands.
$\Delta^0_z$ and $\Delta^1_z$ denote the direct bandgap 
and the energy splitting at $kd_z=\pi$,
respectively.
(c) The variation of $\Delta^0_z$ and 
$\Delta^1_z$ as function of the width ($w_z$) of $N_z$-ZGNRs.
The solid line is a fit curve for $\Delta^0_z$ ($N_z\ge8$)
and the dotted lines are drawn to guide the eyes.
\vspace{-4.0mm}
}
\end{figure}

Upon inclusion of the spin degrees of freedom within LSDA, 
the ZGNR are predicted to have 
a magnetic insulating ground state 
with ferromagnetic ordering at each zigzag edge
and antiparallel spin orientation between the two edges.
The spatial spin distributions of the ground state 
in the case of 12-ZGNR is displayed in Fig. 4(a).
The small spin-orbit coupling ~\cite{dressel2} in carbon atoms 
is neglected in the present study, and we label
one spin orientation as $\alpha$-spin (red)
and the opposite as $\beta$-spin (blue) in Fig. 4 (a).
The total energy difference per edge atom between  
nonspin-polarized and spin-polarized edge states 
increases from 20 meV ($N_z=8$) to 24 meV ($N_z=16$). 
These energy differences are further stabilized 
by an antiferromagnetic coupling between the two edges. 
The total energy difference between ferromagnetic and   
antiferromagnetic couplings between edges, however,
decreases as $N_z$ 
increases and eventually becomes negligible if the width 
is significantly larger than the decay length of the spin polarized edge 
states~\cite{hslee}.
The ferromagnetic-antiferromagnetic 
energy differences per unitcell are 4.0, 
1.8, and 0.4 meV for the 8-, 16-, and 32-ZGNR, respectively. 
Our LSDA results agree with previous 
studies~\cite{fujita,okada,hslee} and consistent with
a theorem based on the Hubbard Hamiltonian
on a bipartite lattice~\cite{lieb}. 
Though infinite range spontaneous magnetic ordering
in a one dimensional Heisenberg model
is ruled out~\cite{mermin}, 
spin orderings in 
nanoscale system are realizable in practice~\cite{gamb,delin,pastor,gamb2} 
at finite temperature assisted by 
the enhanced anisotropy on substrates~\cite{pastor,gamb2}.

The energy gaps in ZGNRs originate from the staggered
sublattice potentials resulting from the magnetic ordering,
which introduce bandgaps for electrons 
on a honeycomb hexagonal lattice~\cite{kane}.
This is realized because 
the opposite spin states on opposite edges
occupy different sublattices respectively 
(black rectangles on the left side 
and grey ones on the right belong to different sublattice respectively
in Fig. 1(b)).
So, the ZGNRs can be considered as 
the magnetic analog of a single BN sheet because 
the former has a bandgap which originates from 
the exchange potential difference on the two sublattices
while the bandgap of the latter is from 
the ionic potential difference between boron and nitrogen atoms 
located on the different sublattices~\cite{blase}.
The Hamiltonian ($\mathcal{H}$) and Bloch wavefunctions
($\psi_{nk\alpha(\beta)}$) of the ground states
satisfy $[{\mathcal H},{\mathcal T \mathcal M}]=0$ and 
${\mathcal T \mathcal M}\psi_{nk\alpha}=\psi_{nk\beta}$
where ${\mathcal T}$
is the time-reversal symmetry operator
and ${\mathcal M}$ a mirror symmetry operator interchanging
sites on opposite sides.
Hence, $\alpha$- and $\beta$-spin states 
are degenerate in all bands and have the same gap as shown 
in Fig. 4(b). 

Since the strength of the staggered potentials in the middle of the ribbon
decreases as the width increases, 
the LSDA bandgaps of ZGNRs are inversely proportional to the width.
The calculated energy gaps (in eV) can be 
fit by $\Delta^0_z(w_z)=9.33/(w_z+15.0)$ 
with $w_z$ in Angstroms when $N_z\ge8$ as shown in Fig. 4(c).
It is also shown that the energy splitting at $ka=\pi$
($\Delta^1_z$ in Fig. 4(b)) converges to 0.53 eV (32-ZGNR) 
from 0.51 eV (8-ZGNR).  

In summary, we have shown that
graphene nanoribbons with homogeneous armchair
or zigzag shaped edges all have energy gaps 
which decrease as the widths of the system increase~\cite{miyake}.
The role of the edges are crucial for
determining the values and scaling rule for the bandgaps.
%We expect that these edge- and width-sensitive energy gaps
%will play an important role in future applications of the GNRs. 

Y.-W.S. thanks F. Giustino, D. Prendergast, and \c C. Girit
for discussions. 
This research was supported 
by NSF Grant No. DMR04-39768 and by
the Director, Office of Science, Office of Basic Energy under
Contract No. DE-AC02-05CH11231. Computational
resources have been provided by NSF at NPACI and
DOE at NERSC.


\vspace{-5.0mm}
\begin{thebibliography}{99}
\bibitem{dressel}G. Dresselhaus, M. S. Dresselhaus, P. C. Eklund, 
               {\it Science of Fullerenes and Carbon Nanotubes : 
                    Their Properties and Applications} (Academic Press, 1996).
\bibitem{louie}L. Chico {\it et al},  
%L. Chico, V.H. Crespi, L.X. Benedict, S.G. Louie, M. L. Cohen, 
%"Pure Carbon Nanoscale Devices: Nanotube Heterojunctions", 
               Phys. Rev. Lett. {\bf 76}, 971 (1996).
\bibitem{mceuen}P. L. McEuen, M. S. Fuhrer, H. Park, 
%               Single-walled carbon nanotube electronics,
               IEEE Trans. Nanotechnol. {\bf 1}, 78 (2002).
\bibitem{fujita}%M. Fujita {\it et al}, 
                M. Fujita, K. Wakabayashi, K. Nakada, K. Kusakabe,
%               Peculiar localized state at zigzag graphite edge,
                J. Phys. Soc. Jpn. {\bf 65}, 1920 (1996).
\bibitem{nakada}%K. Nakada {\it et al}, 
               K. Nakada, M. Fujita, G. Dresselhaus, M. S. Dresselhaus,
%               Edge state in graphene ribbons: Nanometer size effect and
%               edge shape dependence,
               Phys. Rev. B {\bf 54}, 17954 (1996).
\bibitem{waka}%K. Wakabayashi {\it et al}, 
             K. Wakabayashi, M. Fujita, H. Ajiki, M. Sigrist,
%             Electronic and magnetic properties of nanographite ribbons,
             Phys. Rev. B {\bf 59}, 8271 (1999).
\bibitem{ezawa}M. Ezawa,
%              Peculiar width dependence of the electronic properties of 
%              carbon nanoribbons,
              Phys. Rev. B {\bf 73}, 045432 (2006).
\bibitem{brey}L. Brey and H. A. Fertig, 
%              Electronic states of graphene nanoribbons,
              Phys. Rev. B {\bf 73}, 235411 (2006).
\bibitem{sasaki}K.-I. Sasaki, S. Murakami, R. Saito, 
%              Guage field for edge state in graphene,
               J. Phys. Soc. Jpn. {\bf 75}, 074713 (2006).
\bibitem{abanin}D. A. Abanin, P. A. Lee, L. S. Levitov, 
%              Spin-filtered edge states and quantum hall effect in 
%              graphene,
              Phys. Rev. Lett. {\bf 96}, 176803 (2006).
\bibitem{okada}S. Okada and A. Oshiyama, 
%              Magnetic ordering in hexagonally bonded sheets with first-row elements,
              Phys. Rev. Lett. {\bf 87}, 146803 (2001).
\bibitem{hslee}H. Lee {\it et al}, 
              %H. Lee, Y.-W. Son, N. Park, S. Han, J. Yu,
%              Magnetic ordering at the edges of graphitic fragments: Magnetic
%              tail interactions between the edge-localized states,
              Phys. Rev. B {\bf 72}, 174431 (2005).
\bibitem{half}Y.-W. Son, M. L. Cohen, S. G. Louie, Nature, {\it in press} (2006).
\bibitem{miyamoto}Y. Miyamoto, K. Nakada, M. Fujita, 
%                 First-principles study of edge states of H-terminated
%                 graphitic ribbons,
                 Phys. Rev. B {\bf 59}, 9858 (1999).
\bibitem{kawai}%T. Kawai {\it et al}, 
               T. Kawai, Y. Miyamoto, O. Sugino, Y. Koga, 
%              Graphitic ribbons without hydrogen-termination: Electronic structures
%              and stabilities,
               Phys. Rev. B {\bf 62}, R16349 (2000).
\bibitem{hiura}H. Hiura, 
%              Tailoring graphite layers by scanning tunneling microscopy,     
               Appl. Surf. Sci. {\bf 222}, 374 (2004).
\bibitem{novoselov}K. S. Novoselov {\it et al}, 
            %K. S. Novoselov, A. K. Geim, S. V. Morozov, D. Jiang, M. I. Katsnelson, 
            %I. V. Grigorieva, S. V. Dubonos, A. A. Firsov,
%             Two-dimensional gas of massless Dirac fermions in graphene,
             Nature {\bf 438}, 197 (2005). 
\bibitem{kim}%Y. Zhang {\it et al}, 
             Y. Zhang, Y.-W. Tan, H. L. Stormer, P. Kim,
%            Experimental observation of the quantum Hall effect and Berry's
%            phase in graphene,
             Nature {\bf 438}, 201 (2005).
\bibitem{berger}C. Berger {\it et al},
%            Electronic Confinement and Coherence in Patterned Epitaxial Graphene,
            Science {\bf 312}, 1191 (2006).
\bibitem{berger2}C. Berger {\it et al}, 
                  %C. Berger, Z. Song, T. Li, X. Li, A. Y. Ogbazghi, R. Feng, 
                  %Z. Dai, A. N. Marchenkov, E. H. Conrad, P. N. First, W. A. de Heer,
%                  Ultrathin epitaxial graphite: 2D electron gas properties and a route toward
%                  graphene-based nanoelectronics, 
            J. Phys. Chem. B {\bf 108}, 19912 (2004).
\bibitem{sadowski}M. L. Sadowski {\it et al}, cond-mat/0605739.
\bibitem{reich}%S. Reich {\it et al}, 
               S. Reich, J. Maultzsch, C. Thomsen, P. Ordej\'on,
%              Tight-binding description of graphene,
              Phys. Rev. B {\bf 66}, 035412 (2002).
\bibitem{soler}J. M. Soler {\it et al}, 
              %J. M. Soler, E. Artacho, J. D. Gale, A. Garc\'ia, J. Junquera, 
              %P. Ordej\'on, D. S\'anchez-Portal,
%               The SIESTA method for {\it ab intio} order-$N$ materials simulation,
               J. Phys. Condens. Matter {\bf 14}, 2745 (2002)
               and references therein.
\bibitem{gga}We expect that the generalized gradient approximation
             would not change the main results.
% of the present manuscript.
\bibitem{fd}We have tested the various values for electronic temperature
        from 12 K to 580 K but found no variation of the size of the gap.
%\bibitem{troullier}N. Troullier and J. L. Martins, 
%                  Efficient pseudopotentials for plane-wave calculations,
%                   Phys. Rev. B {\bf 43}, 1993 (1991).
%\bibitem{kleinman}L. Kleinman and D. M. Bylander, 
%                 Efficacious form for model pseudopotentials,
%                  Phys. Rev. Lett. {\bf 48}, 1425 (1982).
%\bibitem{kohn}W. Kohn, 
%              Self-interaction correction to density-functional approximations for
%              many-electron systems,
%                Rev. Mod. Phys. {\bf 71}, 1253 (1999).
\bibitem{coulson}C. A. Coulson, 
%             Factors affecting the bond lengths in conjucated and aromatic 
%             molecules,
             J. Phys. Chem. {\bf 56}, 311 (1952).
\bibitem{porezag}D. Porezag {\it et al}, 
%              Construction of tight-binding-like potentials on 
%              the basis of density-functional theory: Application to carbon,
              Phys. Rev. B {\bf 51}, 12947 (1995).
\bibitem{dressel2}G. Dresselhaus, M. Dresselhaus, J. G. Mavrodies, 
%                  Spin-orbit interaction in graphite,
                  Carbon {\bf 4}, 433 (1966).
\bibitem{lieb}E. H. Lieb, 
%              Two theorems on the Hubbard model,
              Phys. Rev. Lett. {\bf 62}, 1201 (1989)
%             Erratum,
%            {\bf 62}, 1927 (1989).
\bibitem{mermin}N. D. Mermin and H. Wagner, 
%               Absence of ferromagnetism or antiferromagnetism in 
%               one- or two-dimensional isotropic Heisenberg models,
                Phys. Rev. Lett. {\bf 17}, 1133 (1966).
\bibitem{gamb}P. Gambardella {\it et al}, 
             %P. Gambardella, A. Dallmeyer, K. Maiti, M. C. Malagoli, 
             %W. Eberhardt, K. Kern, C. Carbone,
%              Ferromagnetism in one-dimensional monatomic metal chains,
              Nature {\bf 416}, 301 (2002).
\bibitem{delin}A. Delin, E. Tosatti, R. Weht, 
%              Magnetism in Atomic-Size Palladium Contacts and Nanowires,
               Phys. Rev. Lett. {\bf 92}, 057201 (2004).
%\bibitem{ying}Y. Li and B.-G. Liu, 
%              Long-range ferromagnetism in one-dimensional monatomic 
%              spin-chain,
%              Phys. Rev. B {\bf 73}, 174418 (2006).
\bibitem{pastor}J. Dorantes-D\'avila and G. M. Pastor,  
%             Magnetic anisotropy of one-dimensional nanostructures of transition metals,
              Phys. Rev. Lett. {\bf 81}, 208 (1998).
\bibitem{gamb2}P. Gambardella {\it et al},
%            Oscillatory magnetic anisotropy in one-dimensional atomic wires,
            Phys. Rev. Lett. {\bf 93}, 077203 (2004).
\bibitem{kane}C. L. Kane and E. J. Mele, 
%            Z$_2$ topological order and the Quantum spin Hall effect,
             Phys. Rev. Lett. {\bf 95}, 146802 (2005).
\bibitem{blase}%X. Blase {\it et al}, 
              X. Blase, A. Rubio, S. G. Louie, M. L. Cohen, 
%             Quasiparticle band structure of bulk hexagonal boron nitride
%             and related systems,
             Phys. Rev. B {\bf 51}, 6868 (1995).
\bibitem{miyake}
LDA Kohn-Sham gaps in general underestimate the quasiparticle band gaps
of semiconductors. 
(See, e.g., M. S. Hybertsen and S. G. Louie, Phys. Rev. B {\bf 34}, 5390 (1986)).
We expect an overall increase in the value of the gap calculated here when considering
quasi-particle corrections using the GW approximation as discussed in
T. Miyake and S. Saito, 
Phys. Rev. B {\bf 68}, 155424 (2003); {\it ibid.} {\bf 72}, 073404 (2005).
\end{thebibliography}
\end{document}